# The fillet of a rock on the Moon: Cohesion and size dependent abrasion rates from topographic diffusion, LRO/NAC and Apollo images

O. Rüsch[1] and C. Wöhler[2]

1. Institut für Planetologie, Westfälische Wilhelms Universität Münster, Münster, Germany
2. Image Analysis Group, Technical University of Dortmund, Dortmund, Germany

Contact details:
Dr. Ottaviano Rüsch
Institut für Planetologie
Westfälische Wilhelms Universität Münster
Wilhelm-Klemm-Straße 10
48149 Münster
Germany
Email: ottaviano.ruesch (at) uni-muenster.de

**Abstract**

The efficiency of regolith production is key in understanding the properties of airless surfaces. Debris aprons, of fillets, around rocks are an ubiquitous morphology on many surfaces without atmosphere, which origin and evolution are largely unknown. Here we show that fillet originates from the juxtaposed rock under abrasion and that rocks of different cohesion have fillets with distinct morphological evolution. Thus, a fillet around a rock allows to disentangle rock cohesion from its surface exposure age. By combing topographic diffusion modeling with images of blocks of known age on the Moon we find abrasion rates for cm-sized boulders similar to regional rates (0.2 mm/Myr), whereas for 10-m sized blocks the rate is two order of magnitude higher (20 mm/Myr). Rates for instances of rocks of higher strength are reduced by ~50%. Fillets around lunar rocks are consistent with abrasion by isotropic micrometeoroid bombardment.

## 1. Introduction

Planetary geology *of proximity* is the area of research that we refer to for investigations of features observable, in theory and in practice, by the naked human eye on a planetary surface, i.e., features from the millimeter to a thousand meter. Up to now, data acquired directly by humans in this spatial domain is limited to the Apollo missions on the Moon. Orbiting spacecraft at low altitude, however, are providing an increasingly higher volume of data in this spatial range. A ubiquitous geological feature of proximity is the block (1-100 m) and the smaller boulder (0.1-1 m) (Bruno and Ruban, 2017). By investigating the evolution of these features after their formation, either by impact ejecta or mass wasting, it is possible to better comprehend how regolith evolves with time, e.g., the type and rate of erosion processes and their interplay with the physical and compositional nature of the surface (e.g., Hörz et al., 1977).

The concept of this article is to gain insights from single instances of this feature (i.e., "*a* rock" as in McDonnell et al., 1977) and not by analyzing an entire population. We choose to concentrate on panchromatic reflected radiation from surfaces in the visible wavelength range. By working with this range, one has access to the highest ground sampling resolution: the resolution being a function of the acquisition time of an orbiting camera and thus controlled by the power of the incoming radiation at the sensor. A drawback of panchromatic light is the challenge of extracting compositional information, a much aimed and critical knowledge. Using the radiation outside the visible wavelengths is the common approach to infer the composition. However, the physical description of the topography and associated modeling is achieved better by harvesting visible images, for example to estimate the age of single craters (e.g., Fassett and Thomson, 2014, Richardson et al., 2020). Here we take an approach similar to that of Fassett and Thomson (2014) to extract age and material strength information from observations of blocks. We focus on blocks of the Moon because of the wealth of information available in terms of their age (in particular surface exposure ages by returned samples) and observations by Apollo astronauts, unmanned landers and Lunar Reconnaissance Orbiter Camera/Narrow Angle Camera (LROC/NAC) (Robinson et al., 2010) images.

The evolution of blocks on the lunar surface is controlled by meteoroid impacts that lead to two morphologically distinct effects: abrasion (e.g., Shoemaker et al., 1970; Gault et al., 1972; Hörz et al., 1974) and shattering (e.g., Hörz et al., 1975a; Hörz, 1977). Shattering, or fragmentation, of lunar blocks was recently described in morphological terms by Ruesch et al. (2020) and literature therein. Here we focus only on erosion by abrasion. This type of process has received relatively little attention in the past due to the fact that erosion by shattering is a more efficient erosive process for blocks (Hörz et al., 1975a). Abrasion, however, is the process responsible for a characteristic lunar morphology, the fillet, an embankment of fine-grained material around rocks (Figure 1) (Shoemaker et al., 1969). A review of the literature reveals that rocks with fillets, sometime identifiable only as "mounds", are ubiquitous features on bodies larger than several km like Phobos (e.g. Thomas et al., 2000), 433 Eros (e.g., Thomas et al. 2001, Dombard et al., 2010), 4 Vesta (Schroeder et al., 2021), 1 Ceres (Schroeder et al., 2021), 21 Lutetia (Küpper et al., 2012), Mercury (Kreslavsky et al., 2021), and potentially on 243 Ida (Fig. 9d in Sullivan et al., 1996). As it was suggested during the Apollo era (Swann et al., 1972), we shall see that the morphometry of the fillet is a useful diagnostic tool. In this work we aim to model the three–dimensional topographic evolution of a block due to abrasion in order to (i) attempt to disentangle temporal and compositional effects, (ii) determine erosion rates of blocks, and (iii) provide simulated images to enable photo-geological interpretation and mapping of blocks under varying stages of erosion. Throughout this study we differentiate between the meaning of the terms "block" and "rock". We use the term "block" to refer to the whole rock-fillet feature. The term "rock" is used to refer to the consolidated material from which the fillet is produced.

## 2. Methods
### 2.1 Modeling of topographic diffusion for a block
#### 2.1.1 The diffusion approximation

Micrometeorite bombardment is responsible for the abrasion of block on the lunar surface (e.g., Shoemaker et al., 1970a; Gault et al., 1972; Hörz et al., 1974, 1977, 2020). Additional processes, such as thermal stresses due to diurnal temperature variations (Molaro and Byrne, 2012) and seismic shaking (Schultz and Gault, 1975) might contribute to their erosion but their importance appears subordinate to meteoroid impacts (e.g., Ruesch et al., 2020). We thus consider micrometeorite bombardment alone in this study. Each micrometeorite impact excavates and melts a very small amount of rock material and deposits a fraction of it around the pit (e.g., Hörz et al., 1975b, Morrison and Clanton, 1979; Loeffler et al., 2008; Harries et al., 2016). This repeated process takes place at a spatial scale much smaller than the rock itself and can be described with the diffusion equation (e.g., Soderblom, 1970; Craddock and Howard, 2000; Howard, 2007). This approach has been used by Fassett and Thompson (2014) to model the topographic evolution of large craters with gentle slopes due to micrometeorite bombardment. Here we follow the same approach for the topography of blocks. As we shall see, the diffusion equation allows us to efficiently model the topographic evolution of a block. The topographically steep edges of a rock represent non-ideal conditions (not a gentle slope) for the applicability of the diffusion approach. However, vertical slopes are only present at the beginning of the simulation period. The diffusion equation requires that the flux of material is proportional to the topographic gradient and that any change in material flux results in an increase or decrease in elevation (e.g., Pelletier, 2008). In one dimension, it can be expressed with the second-order differential equation:

$$\frac{\partial h}{\partial t} = k \frac{\partial^2 h}{\partial x^2} \quad \text{(eq. 1)}$$

with *h* the elevation above the surrounding, *x* the distance and *t* time. As in Fassett and Thompson (2014), we use *k* to denote the diffusivity coefficient. The equation is solved with the Alternating-Direction-Implicit method (e.g., Pelletier, 2008).

**2.1.2 Diffusion applied to blocks**

Proximity observations by the Apollo astronauts have revealed that rocks can partially or completely be surrounded by an embankment of fine-grained material, often in an apron configuration, termed fillet (e.g., Muehlberger et al., 1972). The fillet has an onlap contact with the juxtaposing rock and decrease in thickness away from the rock (Figure 2a). Apollo era literature (e.g., Swann et al., 1971, 1972, Muehlberger et al., 1972) reports that fillets form by accumulation of material abraded away from the adjacent rock and by the ballistic interception of ejecta from distant large impacts due to the rock topography acting as shield. Although we cannot exclude the latter process, it is clear from a visual investigation of blocks of the same age and same nearby context that that the former is the dominant process (Figure 1). In addition, we do not consider rocks near or on sloping terrains that could accumulate loose material on their uphill side (e.g., Swann et al., 1972). Thus, we only consider fillets built up by accumulation of material from the juxtaposing rock. In this context, primary factors responsible for the fillet coverage and morphology were reported to be (i) coherence or friability of the rock, (ii) the original shape of the rock upon reaching its current location, and (iii) the surface exposure age (Swann et al., 1972). Coherence and exposure age are set as variables in the model developed here. To simplify comparisons, the rock shape at the start of the simulation is set constant for all runs with the three axes of a rectangular cuboid a, b and c (a>b>c) defined as b/a=0.8 and c/a=0.54. These ratios are mean values observed for lunar blocks (Demidov and Basilevsky, 2014) and are similar to the shape of fragments produced by impact shattering (e.g., Michikami et al., 2016). In additional runs, topographic variations are introduced in the top face of the rectangular cuboid to assess the role of rock shape irregularities.

The coupling between rock erosion and fillet deposition is performed with two major requirements that are detailed further below: (i) the mass of material eroded from the rock by diffusion is deposited right outside the initial rock boundary and is responsible to build up the fillet deposit; (ii) the fillet deposit that is juxtaposed to the rock is composed of loose particles, i.e., has lower density and higher diffusivity than the rock.

The mass of material eroded from the rock is defined by letting the topography of the rock diffuse (Figure 3). All mass that after diffusion is located behind the original boundary of the rock is classified as eroded. Apollo astronaut images shows that most of the material eroded from the rock is deposited at the rock-fillet boundary (Figure 2b). This observation justifies the relatively simple assumption of material deposition right off the edge of the original rock (Figure 3). In three dimensions, the flow rate of eroded material is not constant around the boundary of the rock. The amount of deposited material at a given location along the rock boundary is thus set proportional to the flow rate at that location. The topography of the rock-fillet system is a combination of the diffused rock and diffused fillet: within the original boundary of the rock the topography is determined by the diffused rock, whereas outside the original boundary of the rock the topography corresponds to the diffused fillet (Figure 3).

**2.1.2 Model with two diffusivity coefficients**

Upon comparison of Apollo astronaut observations with a rock-fillet topography diffused with the same diffusivity coefficient for both fillet and rock sections the mismatch (not shown) between model and observation is evident. The mean slope of the fillet section is observed to be considerably lower than the slope within the abraded rock section. This is the observation that justifies the use of two distinct diffusivity coefficient: one for the rock section ($k_{rock}$) and one for the fillet section ($k_{fillet}$). Because the mean slope of the fillet is lower than the rock section, the coefficient for the fillet section needs to be higher than for the rock. The higher fillet diffusivity is probably related to its unconsolidated nature compared to the solid rock. For consistency, the density of the fillet section is defined as half of that of the rock. The resulting volume increase is used when the eroded mass is deposited right outside the block to build up the fillet.

At a mature erosional stage, Apollo astronaut images show that fillet material onlaps the rock (Figure 2a). This feature is implemented in the numerical code by allowing the rock boundary to move inwards at any location where the fillet becomes higher than the rock. The inward migration stops at a new boundary where the rock has a higher elevation than the fillet. The code solves the diffusion equation with the Alternating-Direction-Implicit method (e.g., Pelletier, 2008) at a time step of 5 kyr and for each point of a 300x300 points spatial grid. Smaller time steps (down to 0.5 ky) have been tested and do not change the topographic results.

In order to match modeled and observed topographic profiles (section 3.2.1) the parameter space of diffusivity coefficients is explored by calculating profiles from rock diffusvity $k_{rock} = 2 \times 10^{-10}$ to $k_{rock} = 6 \times 10^{-4}$ and fillet diffusivity $k_{fillet} = 3 \times 10^{-8}$ to $k_{fillet} = 5 \times 10^{-2}$ (m$^2$/kyr). There are only very limited images that can be used for reliable measurement of the morphometry of rocks with fillet and for which the surface exposure age is known. Although filleted rocks are abundant in Apollo astronaut observations, for reliable slope measurements we used only images where the pointing direction of the camera was perpendicular to a vertical plane across the center of the rock. These instances are listed in Table 1. These topographic profiles were modeled with three constraints: the initial model rock width was set equal to the observed rock width, the initial rock height to width ratio was set to 0.54, and, if available, the exposure age was set to the radiometric exposure age of the rock samples. The parameter $k_{rock}$ and $k_{fillet}$ were set as free parameters. The fitting criteria were (i) the fillet height / rock height ratio, (ii) fillet slope, and (iii) final rock height / width ratio.

## 2.2 Derivation of artificial images

Additional testing of the model was performed using LRO/NAC images. Ideally, the straightforward testing would have seen the comparison of the digital terrain model (DTM) obtained with the method above with the topography extracted from LRO/NAC stereo images. However, production of a block topography from NAC images is hindered by the relatively few pixels resolving it, the high slopes of its sides, and by the strong albedo contrast to the surroundings. We found that the blocks examined in this study are too small to be reconstructed reliably by the NAC-based stereo topographic maps. Shape from shading, a method used to extract small scale features, relies on spatially homogeneous albedo (e.g., Kirk et al., 2003) or a well-characterized albedo pattern, possibly at lower spatial resolution (Grumpe and Wöhler, 2014). As neither of the two holds true for blocks, the shape from shading technique is challenging to apply here. Therefore, instead of extracting topography from NAC images, we used the model topography to produce artificial images and perform the comparison on an image–to–image basis rather than between topographic profiles. Artificial images were produced from the DTM with the following steps:

(i) Calculation of illuminated and shadowed facets according to ray tracing.

(ii) Calculation of the surface reflectance assuming a Lambertian (diffuse) reflectance behavior, where albedo of the block was set to twice the albedo of the regolith and the fillet (in comparison, Bandfield et al. (2011) set the block albedo to 1.5 times the regolith and fillet albedo, but we found that a factor of 2 leads to more realistic rendered images).

(iii) To approximate the properties of the NAC camera system (Robinson et al., 2010; Malin et al., 2007), a point-spread function (PSF) was applied. The PSF was assumed to be of gaussian shape with a full width at half-maximum of 2 pixels in both vertical and horizontal direction.

Artificial images were produced with a wide range of diffusivity values, i.e., $k_{rock}$ ranging from $1\times10^{-5}$ to $6\times10^{-4}$ m$^2$ kyr$^{-1}$ and $k_{fillet}$ ranging from $2\times10^{-4}$ to $1\times10^{-03}$ m$^2$ kyr$^{-1}$.

The LRO/NAC dataset in the area around crater Byrgius A was visually searched for blocks with a major axis of 20±5 m, of roughly cubic shape, and observed at an incidence angle of 55-65º with a ground sampling interval less than 0.7 m/pixel. Where possible, blocks were selected if consistently oriented as in the artificial images with respect to the Sun azimuth and if resting on flat, horizontal terrain without additional topographic features (craters, ridges, smaller blocks). Such criteria allowed meaningful comparison with the blocks modeled in the artificial images. Blocks formed as ejecta by the Byrgius A impact crater were selected for their surface exposure age of 47±14 Myr (Morota et al., 2009), that is sufficiently old for blocks to have developed fillet, and sufficiently young for enough blocks to have survived catastrophic disruption (e.g., Basilevsky et al., 2013). The block major axis of 20±5 m is sufficiently long to recognize morphological properties in NAC images and sufficiently small to have abundant instances in the NAC dataset. NAC images are affected by an artifact that introduces a slight blurring particularly visible between boundaries (Hamm et al., 2016). This so-called "echo" effect occurs only over a very short spatial scale (three pixels) so that it was not necessary to apply a correction to reduce it.

Even if care was taken to select blocks with consistent shapes and without large additional features nearby, the images contain complex scenes, i.e., show nearby tiny rock fragments and small impact craters, and shadows by nearby topography. Thus, the scene complexity and variation hampered an automated procedure and the analysis had to be performed visually. The comparison between the artificial and actual images was based in relative terms and focused on the following morphologies:

- The extent of the fillet away from the rock.
- The shape in planar view of the boundary between fillet and surrounding regolith.
- The type of boundary between fillet and surrounding regolith (e.g, gradual, abrupt).
- The size and shape of the fillet's shadow adjacent to the rock's shadow.

**2.3 Size-dependent diffusivity coefficients**

As shown in studies of large crater diffusion (Fassett et al., 2018), the coefficient of diffusivity is size-dependent. The larger the feature of interest (e.g., crater, block), the larger is the meteoroid projectile that can abrade it without completely shattering it. The benchmark blocks in Table 1 were used to extrapolate the rock and fillet diffusivity coefficients as a function of size.

## 3. Results

### 3.1 Topographic profiles and morphometry

The topographic profiles across the abraded blocks (Figure 4 and 5) shows that independent of the rock age, the height of the fillet at the contact with the rock is determined by the rock coefficient of diffusivity. Considerable changes of the fillet height with time occurs only when the fillet starts to overlap and eventually self-buries the rock. This trend is shown with morphological parameters in Figure 6a. Clearly, the relative height of the fillet at the contact with the rock increases as the rock diffusivity increases, i.e., as the rock strength decreases. This relative height increase of the fillet is not due to a decrease of the rock height. In fact, Figure 6b shows that the rock height to width ratio remains constant for a wide range of rock diffusivities. Only for very high rock diffusivities, i.e. very weak rocks, and after considerable surface residence time (>~100 Myr), the height of the rock decreases considerably.

While the ratio between fillet height and block height remains largely constant with exposure age, the slope of the fillet decreases with time (Figures 4, 5 and 6). The effect played by the size of the initial rock is minimal for sizes in the range 10 cm to 2 m, and start to be significant for friable rocks of size around 10 m (Figure 7).

Visual investigation of the modeled topographic profiles (Figures 4 and 5) reveals that the height distribution as a function of distance from the center of a rock of radius $r_\text{rock}$ can be approximated, albeit roughly, with two exponential functions:

$$h(0 < d \leq r_\text{rock}) = h_\text{rock} \left(\frac{r_\text{rock}}{r_\text{rock}+\Delta d - d}\right)^{-\alpha_\text{rock}} \qquad (\text{eq.2})$$

$$h(d > r_\text{rock}) = h_\text{fillet} \frac{r_\text{rock}}{d^s} \left(\frac{d - r_\text{rock}}{\Delta d}\right)^{-\alpha_\text{fillet}} \qquad (\text{eq.3})$$

with $h$ and $d$ as the height and distance expressed in meters, respectively. The parameter $h_\text{rock}$ is the height of the rock at its center, $h_\text{fillet}$ the height of the fillet at the boundary to the rock, and $\Delta d$ is the distance incremental. For future applications, we provide sets of parameter values that describe a range of morphological stages of abraded blocks (Table 2).

The effects due to an initial topography deviating from that of a rectangular cuboid are shown in Figure 8. Two blocks are simulated with the same irregular base in planar view, i.e., variable angles between vertical faces (>=90 degrees). The first block was set with a constant height (10.8 m) at the beginning of the run (Figure 8a and b). The second block was initialized with irregular elevations (6.4 – 13.0 m) (Figure 8c and d). The initial differences of the top face of the two blocks are still recognizable at 50 Myr (Figure 8a and d). The slope map of the first block shows that the fillet slope is homogeneously distributed with some irregularities on the top right portion (Figure 8b) and identifiable on the second block (Figure 8d). The fillet of the second block is not homogeneously distributed around the rock and is skewed toward the top right. The mean height and slope of the fillet at the boundary to the rock are lower for the second block compared to the first (lower by ~0.5 m and a few degrees).

### 3.2 Calibration of coefficients with images of blocks of known age.
### 3.2.1 Use of proximity optical images

The blocks that were imaged in ideal conditions by Apollo astronauts are shown in the left-hand side panels of Figure 9 and listed in Table 1. The topographic profiles of these blocks were extracted manually and are plotted on the right-hand side panels of Figure 9. On the same diagrams we show the modeled profiles. The topography of Geophone rock was reproduced with modeled morphological parameters within 20% of the actual values (Figure 10) and with the exposure age close to the radiometrically dated sample (Fig. 4a) (Arvidson et al., 1976). The rounded shape of the rock section of Geophone rock is similar between model and observation, although it is not possible to know whether this feature formed by micrometeoroid abrasion or was already present upon block emplacement. Generally, the exact topography is not reproduced because the rocks originally have complex shapes, i.e., not flat topped, that are not taken into account in the modeling. Block "1005" is located close to the rim of Cone crater so that its exposure age could be identified with the age of the crater itself, ~26 Myr (Arvidson et al., 1975). No fragments from block "1005" were sampled by Astronauts to confirm this age, however. The shape of block "1005" was reproduced with the exposure age similar to its inferred age, ~30 Myr. Block "Large mound" was sampled with several different fragments by Apollo astronauts and two radiometric ages are possible: ~50 Myr or 303±18Myr (Stettler et al., 1973; Marti and Lugmair, 1971). The topographic modeling was achieved only with an exposure age of ~300 Myr (Fig. 4e). Block "1006" lies, like block "1005", near the rim of Cone crater so that an exposure age of ~26 Myr could be suggested. Like block "1005", it has not been sampled and there are no radiometric ages for it. Contrary to the case of block "1005", however, it was not possible to reproduce the mound-like topography of block "1006" with an exposure age of 20–30 Myr. Only with a much older exposure age of ~190 Myr the morphological parameters were within 20% of the actual values (Figure 10).

### 3.2.2 Use of orbital optical images

We compared the dataset of artificial images covering a wide range of diffusivity values with actual instances of blocks with visible fillet at the crater rim of Byrgius A (Figure 11). This comparison is performed for specific block properties (size, shape, age) and observational conditions (illumination angle) as described in the method section. Only a handful of instances conformal to the required properties were identified (Figure 11).

#### 3.2.2.1 Fillet diffusivity coefficient

We find that artificial images with fillet diffusivity between $2\times10^{-4}$ $m^2$ $kyr^{-1}$ and $4\times10^{-4}$ $m^2$ $kyr^{-1}$ present the morphological characteristics (Figure 11d, e, f, m, n, o) identifiable in actual images (Figure 11g, h, i, j, k, l). Artificial images with fillet diffusivity lower than $\sim2\times10^{-4}$ ($m^2$ $kyr^{-1}$) have (i) fillets with a sharp boundary with the juxtaposing regolith and (ii) spread away from the mother rock only over a short distance, i.e., less than half the rock minor axis (Figure 11a, b, c). Such configuration of fillet is not observed. Artificial images with fillet diffusivity higher than $\sim4\times10^{-4}$ $m^2$ $kyr^{-1}$ differ from the actual images because (i) there is a smooth transition between fillet and juxtaposing regolith (Figure 11p, q, r) and (ii) once fully buried, the underlying rock is mostly erased and its shape is not recognizeable. The latter characteristic can be seen in Figure 11i and l, where the underlying rock shape is identified with a flat-topped mound. In Figure 11f the square-based rock shape is not recognizeable because of the very high values of $k_{fillet}$ and $k_{rock}$.

#### 3.2.2.2 Rock diffusivity coefficient

With fillet diffusivity of $4\times10^{-4}$ $m^2$ $kyr^{-1}$, we find that the diffusivity values for the rock ($k_{rock}$) can vary between $2\times10^{-5}$ and $<4\times10^{-4}$ $m^2$ $kyr^{-1}$ to reproduce actual morphologies. A rock diffusivity lower than $2\times10^{-5}$ $m^2$ $kyr^{-1}$ does not develop fillet material sufficiently fast to build a fillet relief consistent with the observational dataset, i.e., the fillet has a smooth boundary to the juxtaposing regolith and has a low topographic relief that does not produce a shadow. The upper limit ($<4\times10^{-4}$ $m^2$ $kyr^{-1}$) is defined by one of the assumptions of the model presented in section 2.1.2.

With a fillet diffusivity of $2\times10^{-4}$ $m^2$ $kyr^{-1}$, we find that $k_{rock}$ values in the range $2\times10^{-5}$ to $<2\times10^{-4}$ $m^2$ $kyr^{-1}$ are sufficient to reproduce the actual images of the filleted blocks. Again, $k_{rock}$ values lower than $2\times10^{-5}$ $m^2$ $kyr^{-1}$ produce too little fillet material, and as a consequence the characteristic fillet morphology does not develop. It is to be noted that we focused our survey at Byrgius A on filleted blocks, so that we cannot exclude the presence of few very blocks of high strength that do not have a surrounding fillet. Such blocks would have $k_{rock}$ < $2 \times 10^{-5}$ m$^2$ kyr$^{-1}$. The abundance of this type of block would be low, however, as most blocks at Byrgius A have a spatially resolved fillet.

With the range of values of $k_{rock}$ and $k_{fillet}$ constrained as explained above and valid for 20 m sized blocks, we produced artificial images every 10 Myr of exposure age (Figure 12). After the first ~10 Myr, the morphology of a filleted block observable from orbit is strongly dependent on the $k_{rock}$ value rather than on the surface exposure age.

### 3.2.3 Size-dependent diffusivity coefficients

In the previous section, using Apollo astronaut photographs and LROC/NAC images we modeled and derived diffusivity coefficients ($k_{rock}$, $k_{fillet}$) for a total of seven instances of blocks for which an estimate of the surface exposure age is available (Table 1). These instances cover a range of block widths from 2 to 20 m. The relation between diffusivity coefficients and block size is assumed to follow a power law, similar to the size-dependence of the diffusivity of craters proposed by Fassett et al. (2018). With such assumption the relation can be derived as follow:

$$k_{rock\_high\_strength} = 3 \times 10^{-8} \, a^{2.16} \quad \text{(eq. 4)}$$
$$k_{rock\_low\_strength} = 6 \times 10^{-7} \, a^{1.7} \quad \text{(eq. 5)}$$
$$k_{fillet} = 2 \times 10^{-6} \, a^{1.83} \quad \text{(eq. 6)}$$

with $a$ as the major axis of the block. For $k_{rock\_high\_strength}$ the coefficients calculated for Geophone Rock, Apollo 17 (Figure 9a) and blocks at Byrgius A (Figure 11) were used. For $k_{rock\_low\_strength}$ the coefficients were taken from Block 1005, near B3 station, Apollo 14 (Figure 9b) and from blocks at Byrgius A (Figure 11n). For $k_{fillet}$ the coefficients are from Geophone Rock, Block 1005 and a block at Byrgius A (Figure 11m, n). These relationships are representative for lunar rocks although they are not endmembers. This means that more cohesive and more friable rocks likely exist on the lunar surface than the two cases presented here.

### 3.3 Abrasion rates

With equations (2)-(4) we model the diffusion of blocks for a range of block sizes and calculate their abrasion rate for the first 100 My of surface exposure. The abrasion rate is defined as the volume of rock material lost after diffusion for a given exposed area and a given exposure period. The abrasion rates as a function of surface exposure age and size are presented in Figure 12 and reported in Table 3. Clearly, the abrasion rate is neither uniform in time nor in space, i.e., diffusion is stronger on the edges of the rock. Figure 12 shows that the time needed for self-burial is inversely proportional to the size because the abrasion rate is higher for larger blocks.

## 4. Discussion
### 4.1 Morphological parameters

The distinct variation with time of the ratio fillet height / block height and of the fillet slope allows to disentangle the effect of rock strength (diffusivity) from the surface exposure age. Theoretically, for given block dimensions (three major axes), simple morphological parameters (Figure 6) are sufficient to provide information on block strength and residence time.

Demidov and Basilevsky (2014) reported that the height to maximum diameter ratio of small rocks in the range 10 cm to 2 m observable in Apollo and Lunokhod-1 and 2 images is relatively uniform across many sites with a mean value of 0.54 ± 0.03. This observation is explained by the fact that micrometeoroid abrasion and possibly other processes are not able to change sufficiently the shapes of rocks (Demidov and Basilevsky, 2014) before catastrophic disruption of the rocks occurs by a larger meteoroid hit (Hörz et al., 1975a; 1977; McDonnell et al., 1977). Our modeling results (Figure 6b) support this interpretation.

Irregularities on the top right portion of the fillet of the two blocks are possibly due to the large angle between two faces in planar view (Figure 8b and d). It can be seen that the fillet deposit of the second block was controlled by the topography of the rock. Considerable fillet material was deposited at the termini of a gentle sloping area of the rock. The fillet closest to the highest elevation ("peak") of the rock (orange in Figure 8c), instead, did not grow and diffuse as much. Overall, topographic differences at the beginning of the simulations lead to minor differences in fillet slope and height compared to the range produced by variations in the coefficients of diffusivity and exposure time (Figure 6).

The results indicate that the two mound-like blocks (Large mound and block "1006") require an old exposure age for their topography to be reproduced. Such mounds could represent rare blocks that never experienced shattering by a large meteoroid and which topography has been controlled only by abrasion. It remains to be explained why at the Apollo 12 landing site two such mounds were observed very close to each other ("Large mound" and "Small mound",

Shoemaker et al., 1970b). It is possible that both two nearby blocks survived catastrophic shattering for a long period. Alternatively, their close proximity might suggest a compositionally controlled evolution, i.e., they might represent young blocks of very friable material. Our findings support the former option.

**4.2 Abrasion rates and compositionally dependent block evolution**

Abrasion rates have been reported in the literature from estimates of returned samples as well as from modeling efforts. Our abrasion rate estimates (Table 3, Figure 13) for low-strength rock (0.2 mm/My) is similar to previous numerical modeling (0.4 mm/My) as well as with analysis of returned Apollo samples for a 10 cm rock (0.3 mm/My, Crozaz et al., 1971). Our estimate for high-strength rock (0.07 mm/My) is within the range of previous modeling (Hörz et al., 1975; Ashworth and McDonnell, 1973), although it is lower than the minimum estimate from Apollo samples for a 10 cm rock (Table 3). We note that there is a great variability of abrasion rates reported in the literature (an order of magnitude) that can be attributed to different techniques and the inherently variable strength of rocks. The fact that our upper estimate falls within the range of other studies and of returned samples confirms the validity of our rates. One reason for the difference between our lower estimate (high-strength rock) and the Apollo samples is that our estimate is a mean for the entire rock after 100 My whereas the Apollo measurements were taken on discrete sections of the returned samples. In addition, the higher rates for the returned samples might be due to their shorter surface exposure age with respect to our 100 My calculation. We find an abrasion rate in the range 0.4-0.9 mm/My (hard and weak rock) in the first 10 My for a rock 10 cm in size. A section of sample 12063 of similar size has been reported to have an abrasion rate of 0.7 mm/My with an exposure age of <1.6 My (Crozaz et al., 1971). We therefore conclude that there is reasonable agreement between the abrasion rates we report and previous estimations.

The dependency of the topography on the rock coefficient of diffusivity ($k_{rock}$) exemplified by the evolution of block morphologies as observable from orbit (Figure 12) demonstrates that rock strength plays a major role in the efficiency of micrometeoroid abrasion. Basilevsky et al. (2013) noted the same dependency on material strength for the case of catastrophic shattering and rock abundances. Therefore, because both abrasion and shattering are strongly dependent on rock material strength, rock abundances cannot always be interpreted in terms of surface exposure age alone. In the first 100–200 Myr, before most (99%) rocks are destroyed by catastrophic shattering, both exposure age and material strength should be considered to explain a certain population of blocks. In the light of this dependence on material strength, we note that the generally higher abundance of blocks and thinner regolith on the maria relative to the highlands could be partially explained by a difference in material strength, and not only by an age difference.

O'Brien and Byrne (2021) estimated a coefficient of diffusivity for regolith of 0.1 m$^2$/Myr for a surface area less than 8 m, i.e., subject to impact craters smaller than 8 m. This value is very close to the diffusivity of 0.09 m$^2$/Myr for a feature of 8 m in size calculated with eq. (6) presented above.

Contrary to the coefficient of diffusivity, abrasion rates do not always increase with increasing spatial scale because abrasion rates are a function of the coefficient of diffusivity and initial topographic gradient. As suggested by Fassett and Thompson (2014), we find that at the local scale the surface is more dynamic than at regional scale. For example, a 1 m and a 10 m wide block experience degradation at a rate of 0.8 mm/Myr and 9 mm/Myr, respectively. The mean erosion rate at large scale (kms wide area) has been estimated at 0.4 mm/Myr (Fassett and Thomson, 2014) and at 0.2 mm/Myr (Craddock and Howard, 2000), similar to the rate of cm-sized rocks.

The rates reported for 1 m and 10 m wide blocks, i.e., 0.8 mm/Myr and 9 mm/Myr, refers to rock material of relatively high strength. These rates increase by a factor 2 for rocks of lower strength (higher $k_{rock}$).

**4.3 Self-burial**

The graph of Figure 12 suggests that about 600 My are needed for blocks to be completely eroded by abrasion only. Before complete abrasion occurs, however, the modeled profiles (Figure 4) show that self-burial occurs by fillet onlap. This means that if catastrophic disruption does not occur, the next major morphological event that happens to a rock is self-burial rather than complete abrasion. Self-burial by the fillet occurs between 100 and 200 My for a 5 m sized rock. After self-burial, the layer of fine particles above the rock can act as a shield from micrometeoroids. Thus, unless the protective layer of fines is not disrupted by a sufficiently large micrometeoroid, the rock will not be subject to further erosion. In general, the presence of intact rocks within the regolith (e.g., Thompson et al., 1974; Ghent et al., 2016; Wei et al., 2020) can be explained by burial from nearby crater ejecta blankets. The self-burial mechanism presented here could be an additional process to explain the presence of subsurface rocks and the complex history of the Apollo samples involving multiple episodes of surface exposure, flipping and burial (e.g., Fleischer et al., 1971). We note that the reverse process can be observed on the central peak of large craters (or other exhumation area) where regolith cover is slowly removed from rocks (e.g., Greenhagen et al., 2016).

**4.4 Model limitations**

Although the model developed here allows to understand the morphologies of lunar blocks and boulders, the comparison of the model outputs, i.e., profiles and artificial images, with actual observations reveals the complex shapes of actual lunar rocks. The most frequent shape of blocks can be approximated by a rectangular cuboid, however, variation exists in the values of the three major axes. For example, the c/a ratio has been reported to vary between 0.49 and 0.58 (Demidov and Basilevsky, 2014). Added complexity occurs where the fillet starts to cover part of the rock. Apollo reports show that the fillet material can be deposited on top of the undulating rock shape well before it reaches the top of the rock (Swann et al., 1972). In this condition, part of the rock would be protected from abrasion, and as a consequence the diffusion of the rock-fillet system would be modified.

**5. Conclusions**

We investigated the morphology and topography of single instances of blocks on the lunar surface. We concentrated on blocks whose erosion history has been controlled by micrometeoroid abrasion without shattering. Morphologically, these blocks consist of the actual rock component surrounded by a particulate depositional apron, i.e., the fillet. The outcomes of this study can be summarized as follows:

- Modeling of lunar block abrasion by diffusion with two coefficients of diffusivity for the rock and fillet sections is sufficient to reproduce the main morphometrical characteristics of blocks, namely the height of the fillet with respect to the height of the rock and the fillet slope.
- The height of a fillet at the contact with the rock is proportional to the rock friability, whereas the slope of the fillet is controlled mainly by the surface exposure age, with a reduction in steepness with time. In low resolution orbital images of airless surfaces, rocks with fillets can be identified by a bright halo around a rock and by the fillet shadows. A map of such features would provide geographical information on the petrology and exposure age of blocks, for example around large impact craters.
- For the first time, lunar abrasion rates are provided as a function of rock size, rock diffusivity (proxy for rock strength) and surface exposure time. These rates were derived by coupling the diffusion model to the morphometry of blocks for which the age is known. The new abrasion rates are in agreement with previous estimates and Apollo samples. Rock abrasion rates increase with rock size. For cm-sized rocks the rates are similar to regional scale erosion rates (0.2 mm/Myr), whereas for 10 m-sized rocks the rates are two order of magnitude higher.

- The topography of a block, and thus its abrasion history, are strongly dependent on rock strength. For rocks of the same size, same age and different strength, the abrasion rate can vary by at least 50 %. Therefore, lunar rock abundances should be interpreted in terms of surface exposure age as well as material strength.
- Fillet development leads to self-burial of rocks. Self-burial is reached before complete abrasion of the rock by micrometeoroids alone. On the average, however, catastrophic disruption by a relatively large meteoroid occurs before self-burial. Where catastrophic disruption does not occur, self-burial is an additional mechanism to explain the presence of subsurface rocks within the regolith.
- In future lunar exploration activities, time-of-flight cameras that allow to reconstruct the three-dimensional shape of blocks independent of the illumination conditions might be used for surface proximity observations to estimate the age and rheology/mineralogy of blocks and smaller boulders.

**6. Acknowledgment**

O.R. is financially supported by a Sofja Kovalevskaja Award of the Alexander von Humboldt foundation. The authors are grateful to Kay Wohlfarth for providing his raytracing code used in the construction of the artificial images in Figures 11 and 12. Discussion of the Apollo 12 mounds geology with Wajiha Iqbal is acknowledged.

**7. Author contributions**

O.R. conceived the study, developed the numerical model and wrote the manuscript draft. C.W. contributed to the development of the concepts, produced the artificial images and edited the manuscript. Both authors discussed and interpreted the results.

**8. Data availability**

LROC/NAC images are available through the Planetary Data Systems Geosciences node (https://pds-geos-ciences.wustl.edu/missions/lro/), and the LROC website

(https://www.lroc.asu.edu/archive). Apollo images used in this study are available at the Apollo Lunar Surface Journal website (https://history.nasa.gov/alsj/main.html).

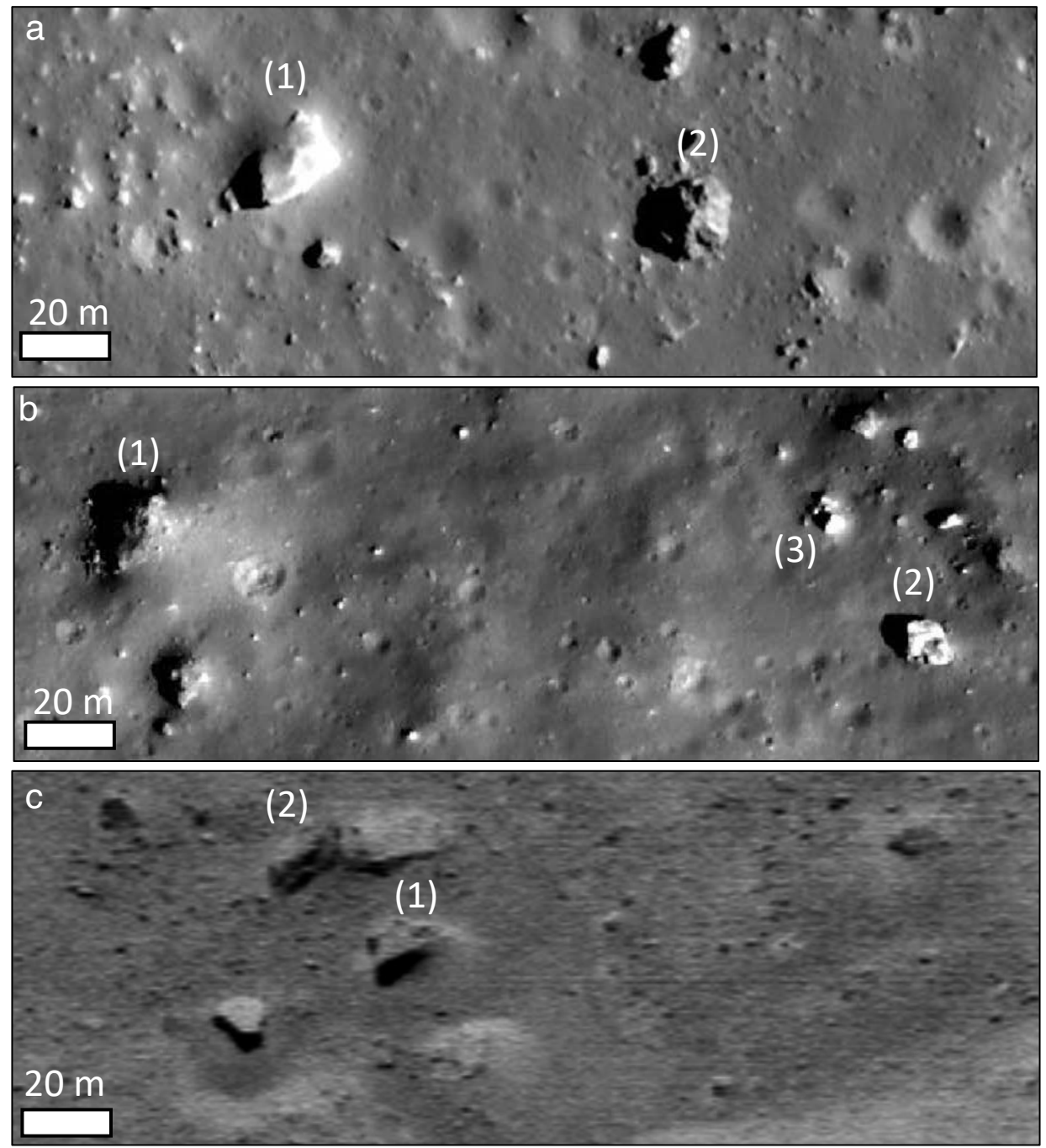

**Figure 1.** (a, b) Examples of distinct morphologies for lunar blocks of approximately the same size and the same surface exposure age. (a) Blocks near the rim of North Ray Crater exposed for 50 My, NAC image M175179080LR. (b) Blocks near the rim of Byrgius A crater, exposed for 47±14 My, NAC image M1129909561RC. In both images, the blocks labeled (1) and (2) have about the same size. Blocks labeled (1) are surrounded and partially embedded by unconsolidated material (fillet). The blocks labeled (2) display no visible fillet. In (b), the cluster of rocks labelled (3) is likely a shattered block as described in Ruesch et al. (2020). (c) Example of fillet morphologies of different development stages around blocks on near-Earth asteroid (433) Eros. NEAR mission image M0157415638F3.

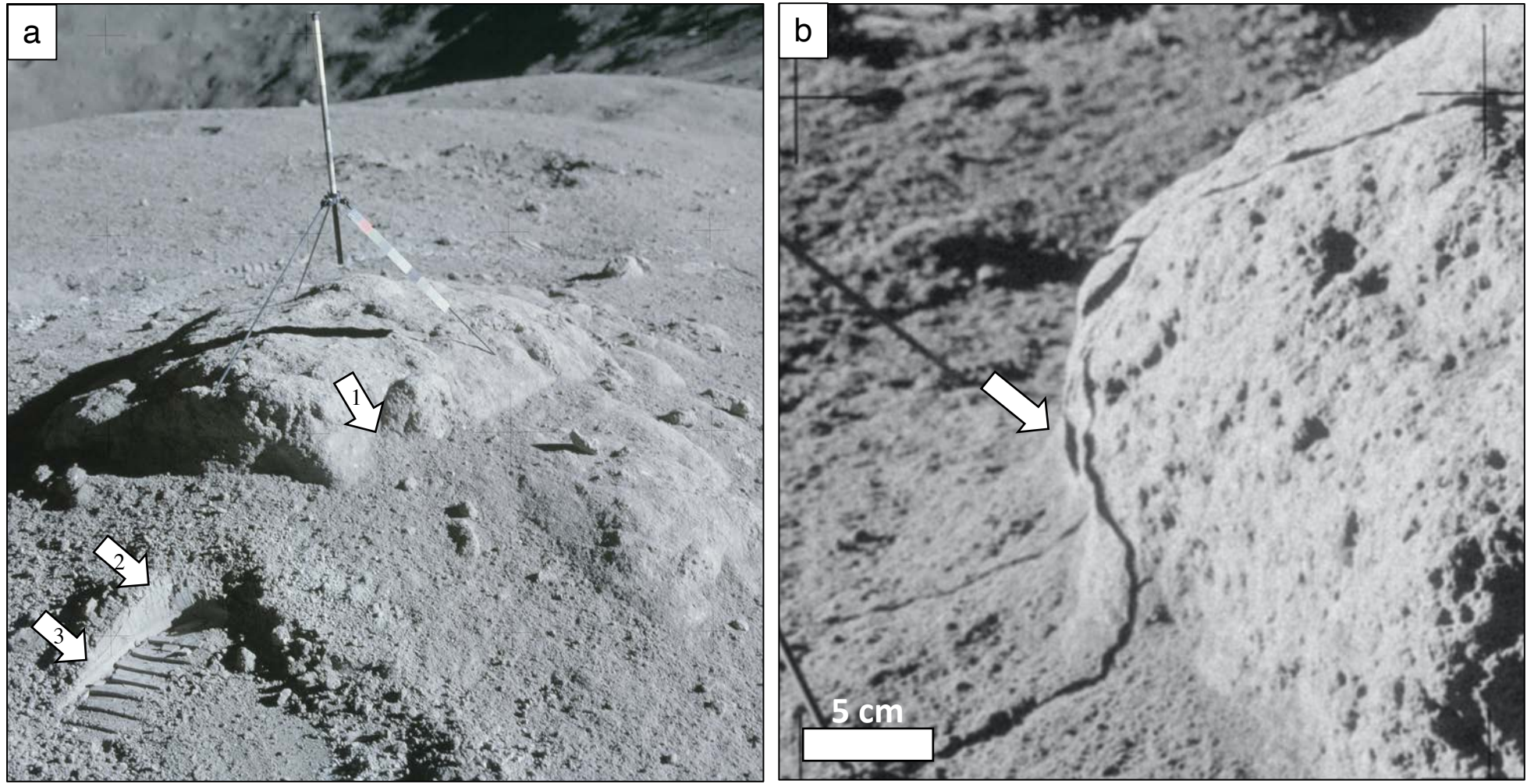

**Figure 2.** Examples of the morphological properties of fillets around rocks with rounded edges. In (a), arrow 1 points to the onlap contact of the fillet onto the rock. Arrows 2 and 3 point to the decreasing thickness of the fillet away from the rock, exposed by an astronaut footprint. Ejecta of nearby crater Plum may have contributed to fillet accumulation. Entire height of gnomon is 62 cm. In (b), the arrow points to the accumulation of fillet right at the contact with the rock. (a) AS16-114-18413. (b) AS12-48-7062.

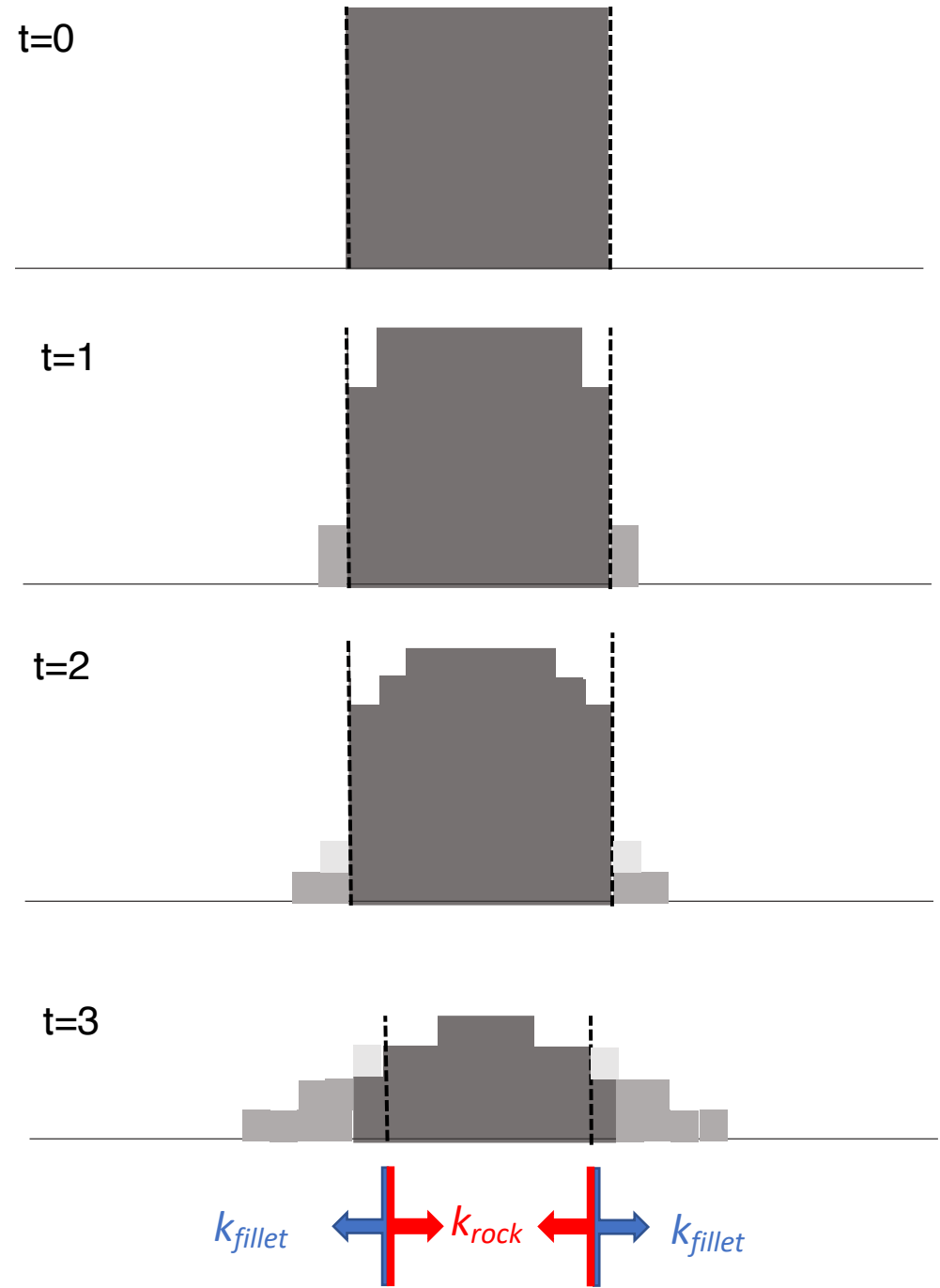

**Figure 3.** Vertical section of a modelled block illustrating how erosion is simulated. An initial block topography (here a square in two dimensions) has a boundary indicated by dashed lines. The block material is subject to diffusion with a diffusion coefficient $k_{block}$. The block material that diffuses behind the boundary is considered to represent unconsolidated fillet material. Fillet material is deposited right outside the boundary and is allowed to diffuse with a diffusion coefficient $k_{fillet}$. When and where the fillet material has the same height as the block at the boundary, the boundary is shifted inwards at the subsequent iteration: this enables the fillet to onlap the block.

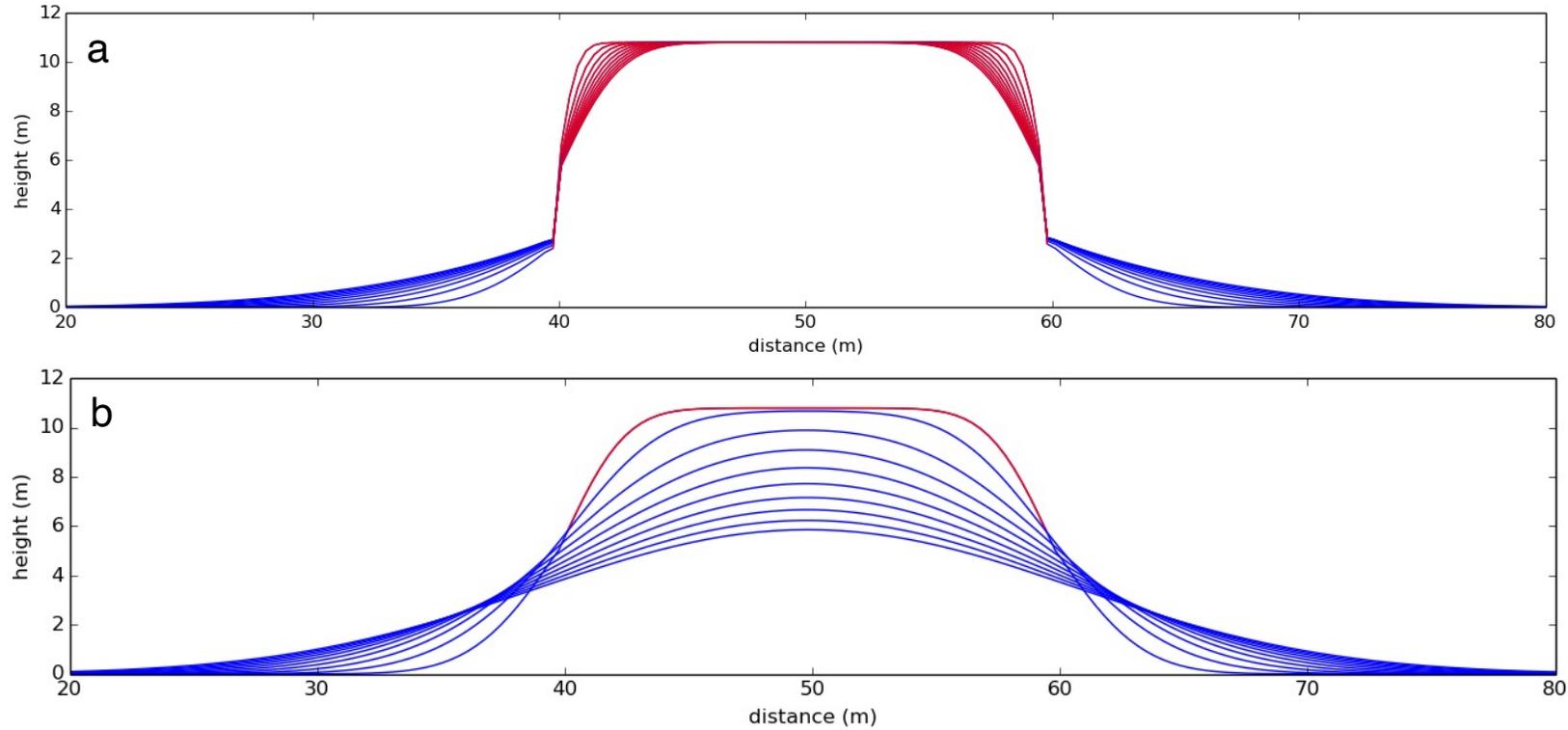

**Figure 4.** Models of topographic profiles for rocks with surface exposure ages from 10 My to 100 My. Red and blue lines indicate block and fillet profiles, respectively. (a) Block of $k_{rock}$=2.0E-5 m²/ky. (b) Block of $k_{rock}$=2.0E-4 m²/ky. Both blocks have fillet diffusivity $k_{fillet}$=4.0E-4 m²/kr, initial width of 20 m and initial height/width ratio of 0.54.

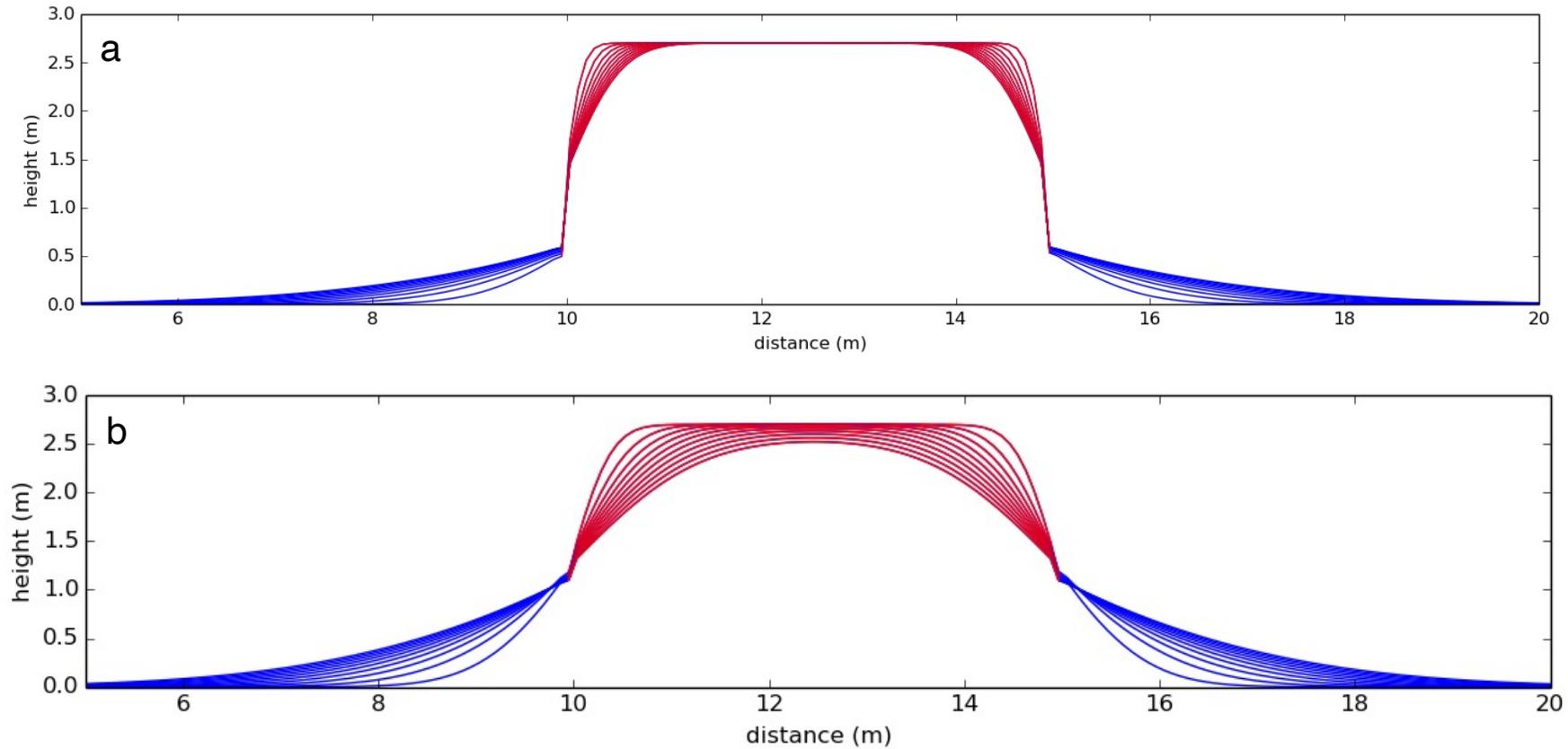

**Figure 5.** Same as Figure 4 for 5 m blocks. Models of topographic profiles for rocks with surface exposure ages from 10 My to 100 My. Red and blue lines indicate actual block and fillet profiles, respectively. (a) Block of $k_{rock}$=1.0E-6 m²/ky. (b) Block of $k_{rock}$= 5.0E-6 m²/ky. Both blocks have fillet diffusivity $k_{fillet}$=3.0E-5 m²/kr, initial width of 5 m and initial height/width ratio of 0.54.

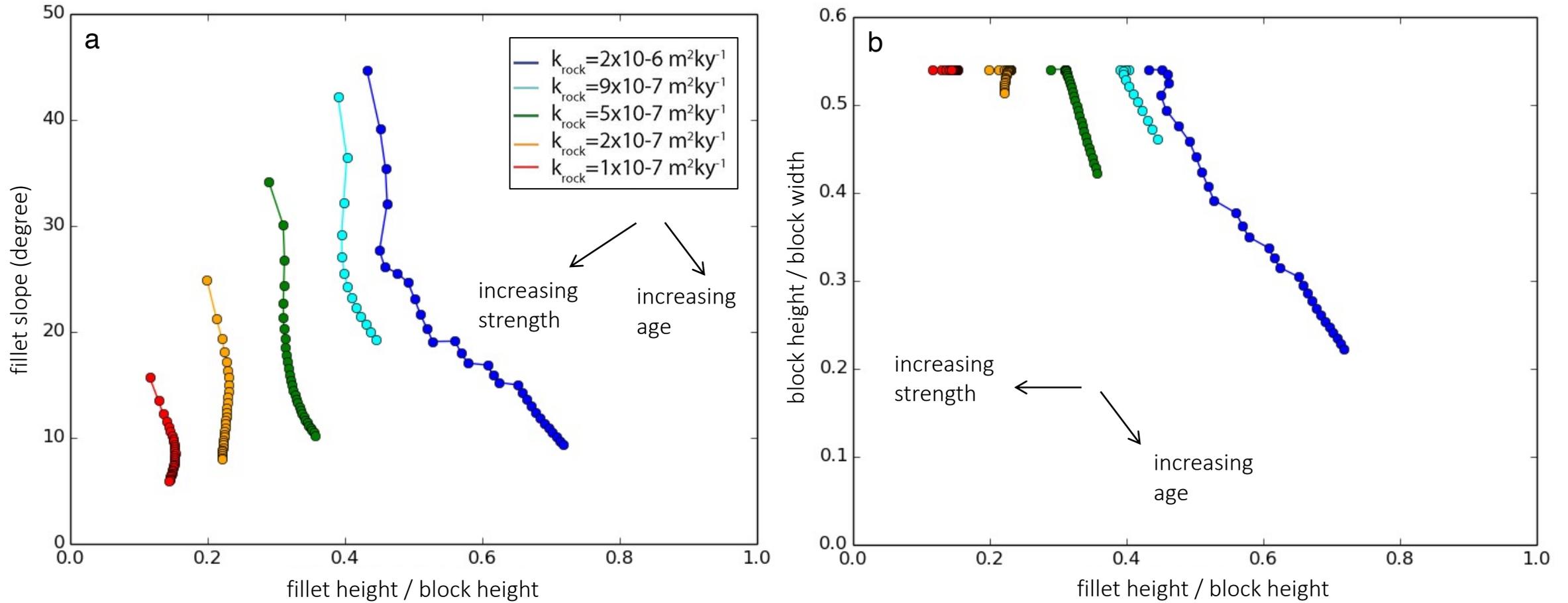

**Figure 6**. (**a**) Morphological parameters measured on modeled blocks 2 m in size with $k_{fillet}$=6x10-6 m$^2$ ky$^{-1}$. The simulated tracks start at the top at a surface exposure age of 10 Myr, the time interval between subsequent points is 10 Myr. Trends due to increasing exposure age and increasing strength (different color) can be disentangled. (**b**) Same models as in (a) with different morphological parameters showing how the height of a block is largely insensitive to diffusion (abrasion) expect for very weak rock material (high $k_{rock}$).

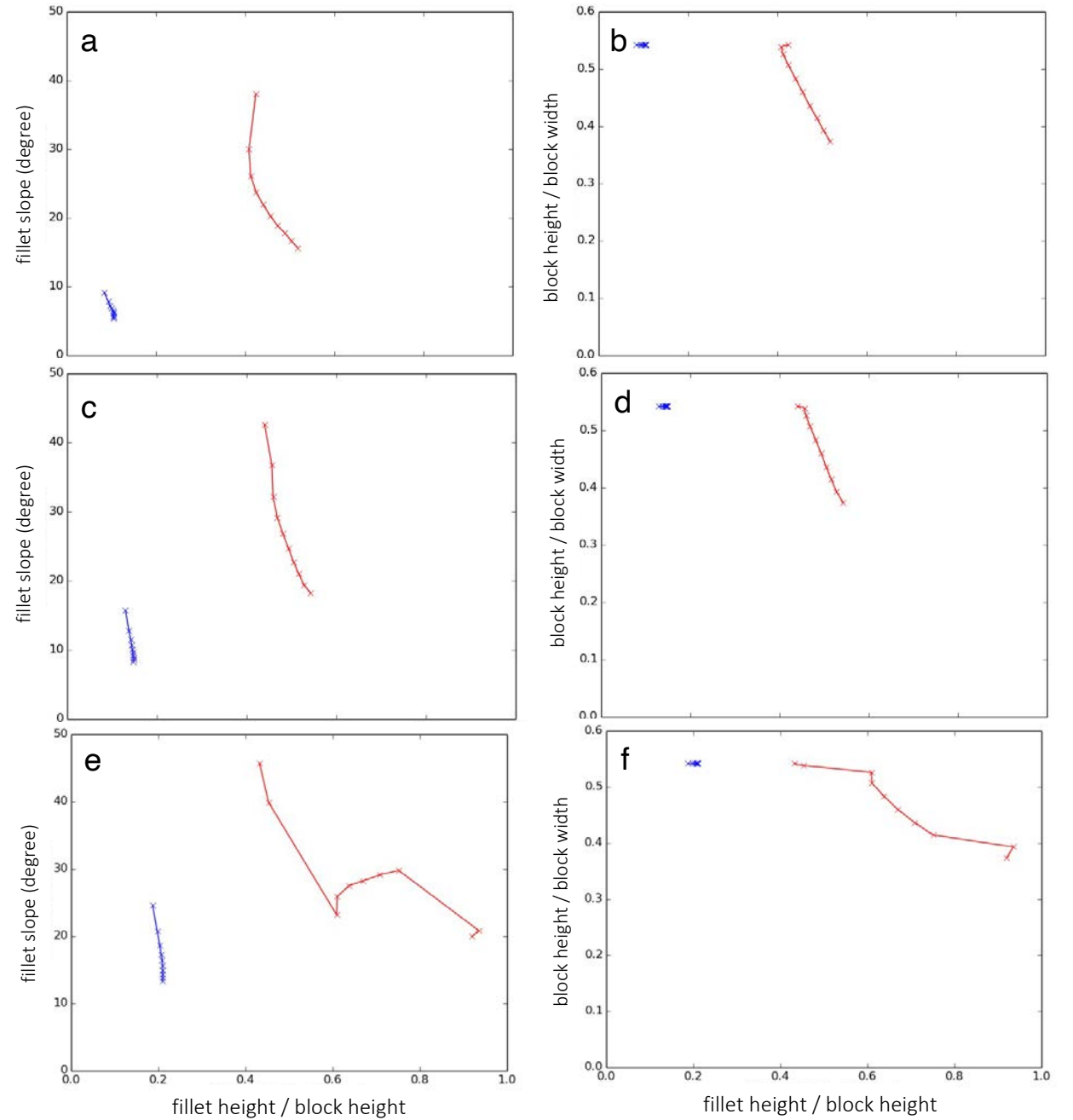

**Figure 7**. Morphological parameters as in Figure 6 for different initial rock sizes. (a,b) 10 cm wide rocks. (c,d) 1 m wide rocks. (e,f) 10 m wide rocks. Blue and red lines denotes high and low strength rocks, respectively. The simulated tracks are 100 Myr long and start at the top at an age of 10 Myr. The time interval between subsequent points is 10 Myr.

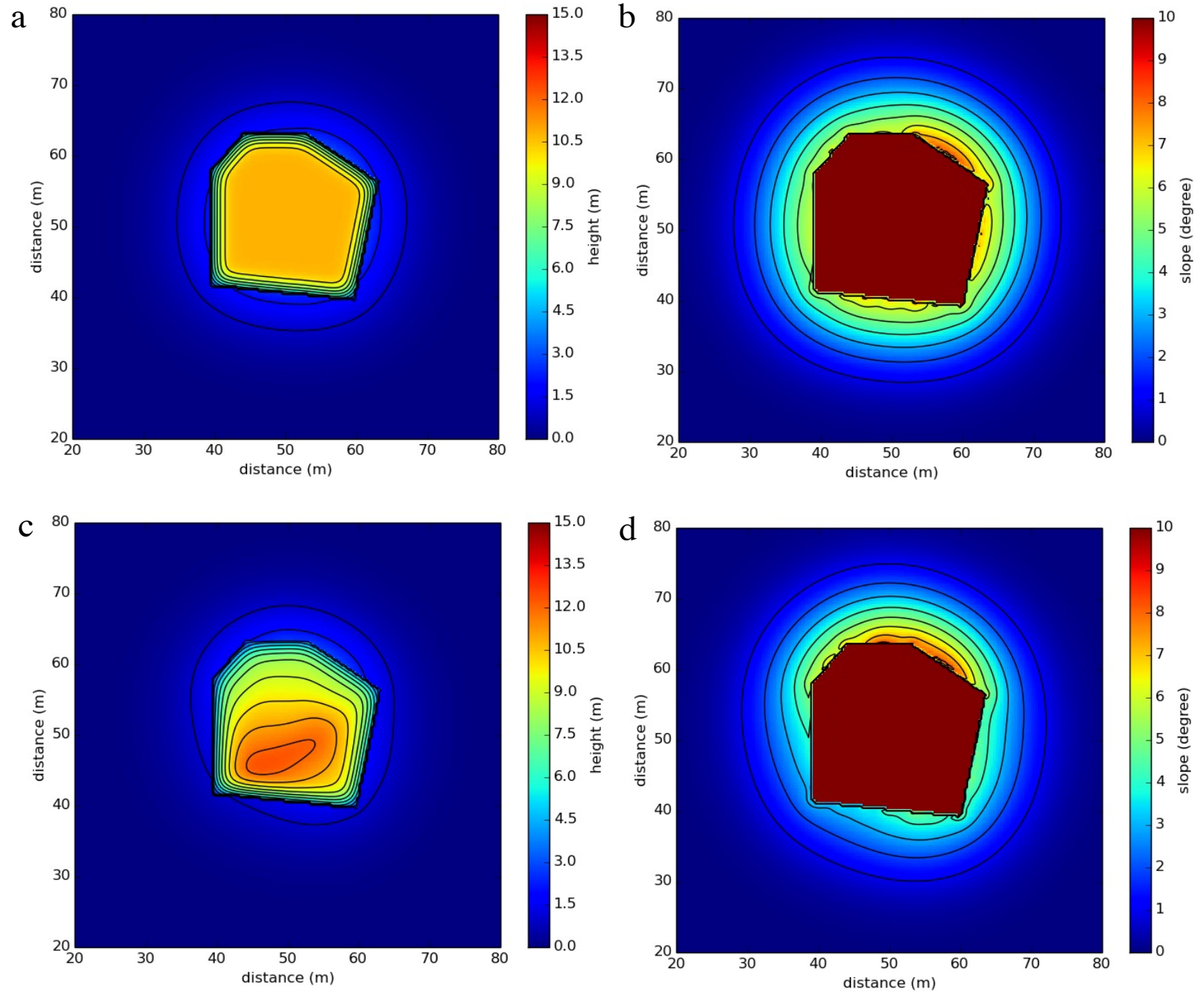

**Figure 8.** Effects of an uneven top face of a block. (a, b) Elevation and slope maps for a block with an initially flat top side. (c, d) Elevation and slope maps of a block with initially irregular heights. Blocks have the same properties (50 My age, $k_{rock}$=1.94E-05 m²/ky, $k_{fillet}$=4.77E-04 m²/ky) aside from the initial top face topography. Contours lines are each m in elevation maps and each degree in slope maps. For clarity, the slope within the rock section is not shown in b, d.

**Figure 9.** Comparison of topographic profiles between observed and modelled blocks. (a,c,e,g) are Apollo astronaut images. (b,d,f,h) are the corresponding topographic profiles (in black) and models (in red). The blocks in these examples are of different sizes and surface exposure ages (Table 1). Model parameters are given in Table 1. (a) AS17-147-22561HR, (c) AS14-68-9432HR, (e) AS12-46-6793HR, (g) AS14-68-9432HR.

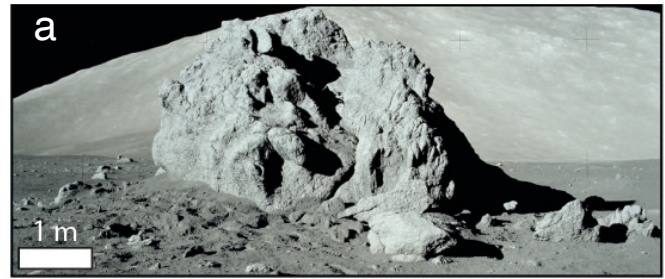
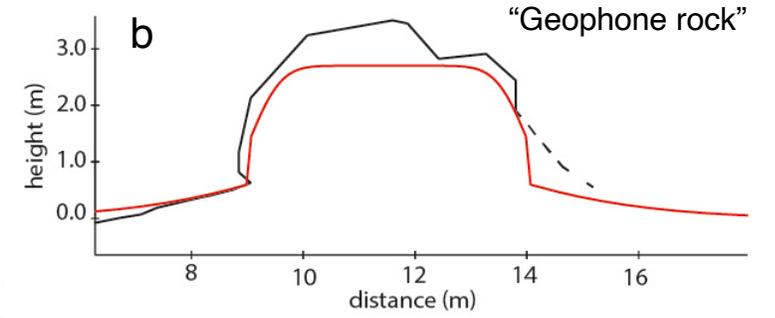
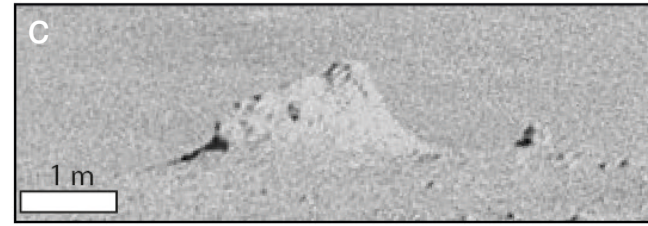
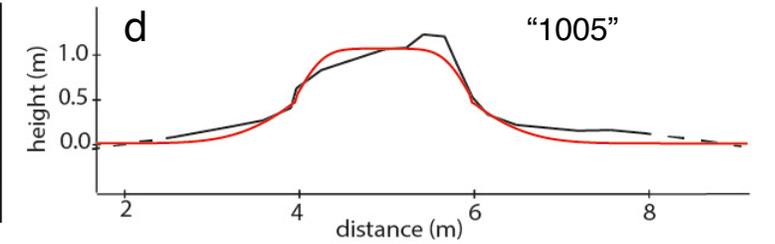
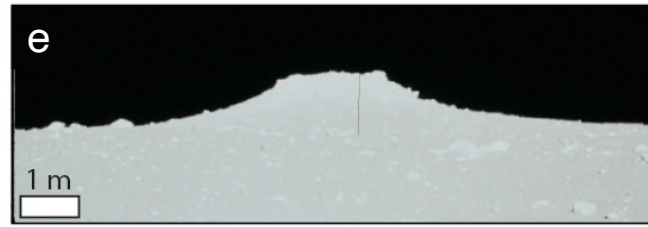
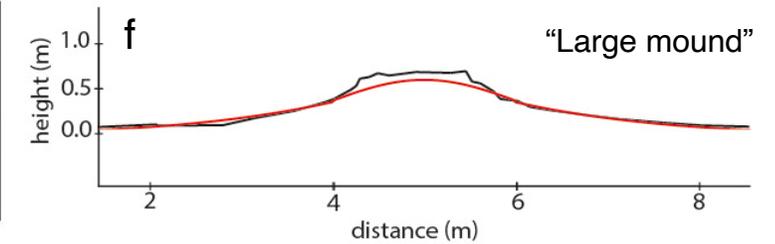
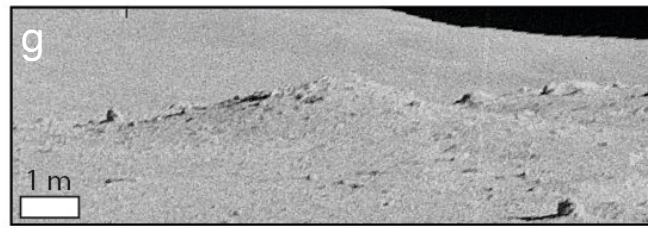
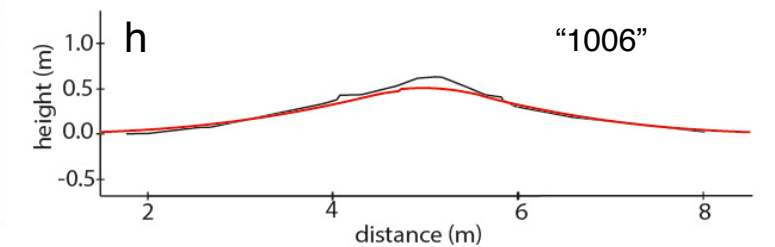

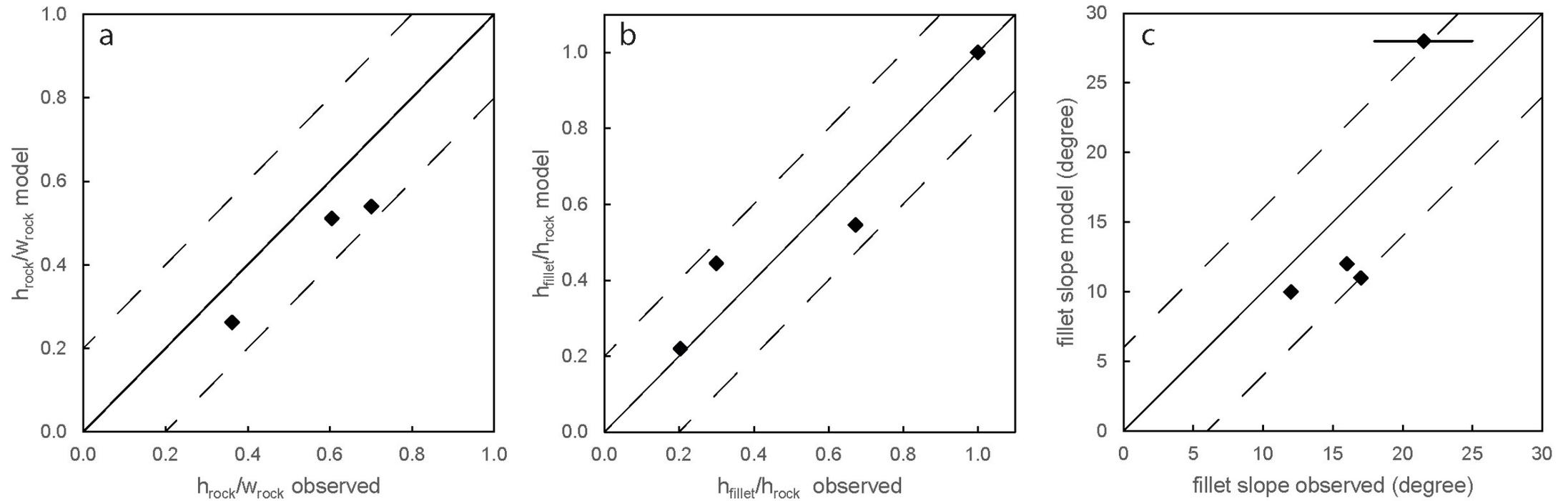

**Figure 10.** Scatterplots of parameter values from observed and modeled topographic profiles shown in Figure 8. (a) Ratio of height of rock to width of rock. (b) Ratio of maximum height of fillet to maximum height of rock. (c) Fillet slope calculated for the first third of the fillet profile starting from the rock-fillet boundary. The data point with uncertainty bar represents block "1005" with an asymmetric profile. The dashed lines represent the 20% deviation from the bisection line.

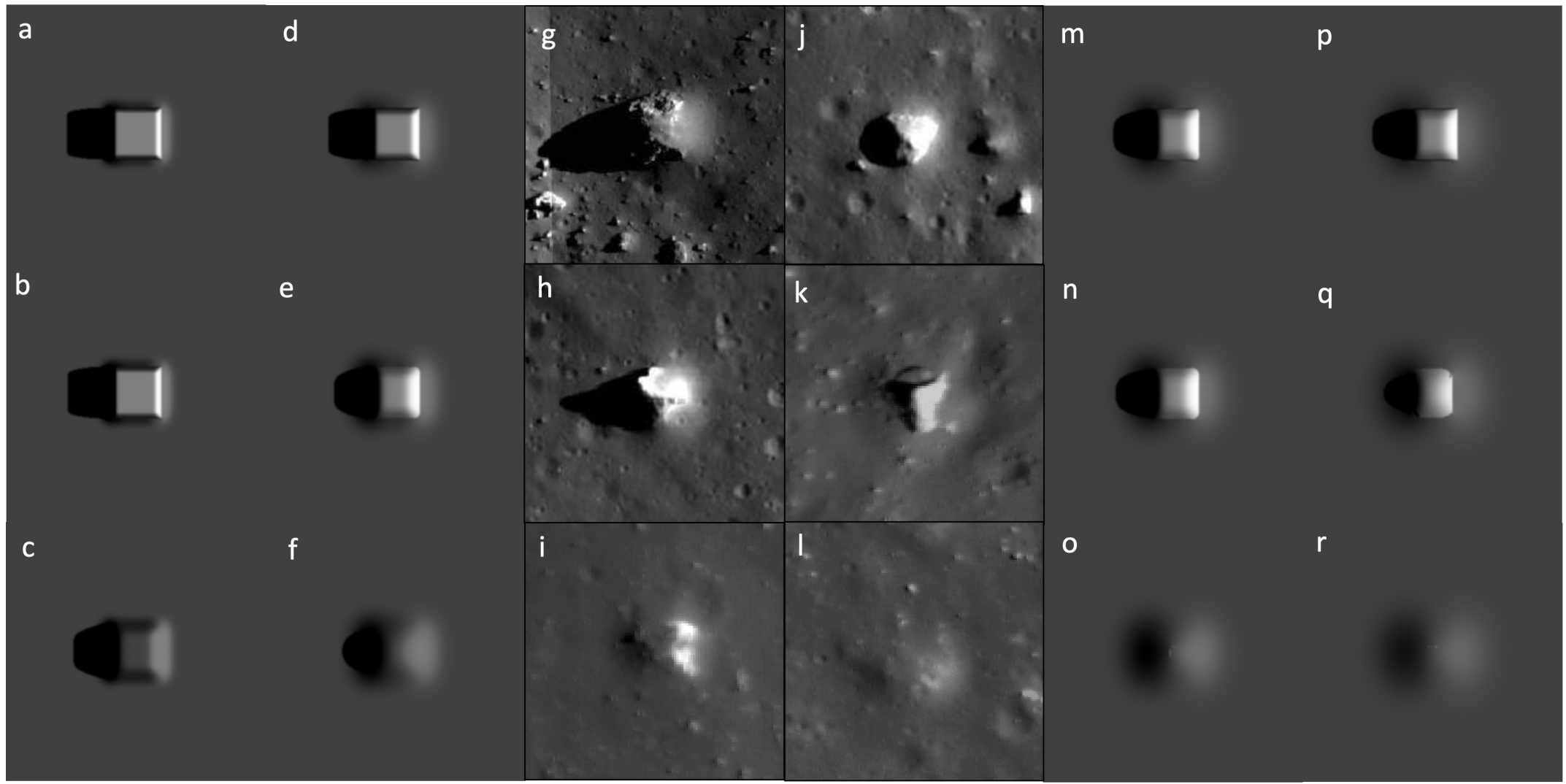

**Figure 11**: Comparison between LROC/NAC observations at Byrgius A crater rim (g,h,i,j,k,l) and simulated images (a,b,c,d,e,f,m,n,o,p,q,r). The coefficient of diffusivity of the fillet increases in the artificial images from left to right. The coefficient of diffusivity for the rock increases from top to bottom. The morphology of the actual fillets and rocks (g,h,i,j,k,l) is reproduced by panels (d,e,f) and (m,n,o). The panels (a,b,c) and (p,q,r) have slightly different morphologies, e.g., sharpness and shape of the fillet-regolith boundary, extension of fillet away from the rock. Observed blocks are 20-3+6 m in width and ~47±14 Myr in age. Modelled blocks are 20 m in width and 50 Myr in age. (a,b,c): $k_{fillet}$ =5E-05 (m²/kyr), (d,e,f): $k_{fillet}$ = 2.0E-04 (m²/kyr). (g,h,i): Byrgius A. (j,k,l): Brygius A, (m,n,o): $k_{fillet}$ = 4.0E-04 (m²/kyr). (p,q,r): $k_{fillet}$ = 6.0E-04 (m²/kyr). Incidence angle is at 60° in artificial images and in the range 54°–67° in the NAC images. g,j,h: NAC image M145047737L. k: NAC image M1354540887R i, l: NAC image M1265236990R. Illumination is from the right in all images.

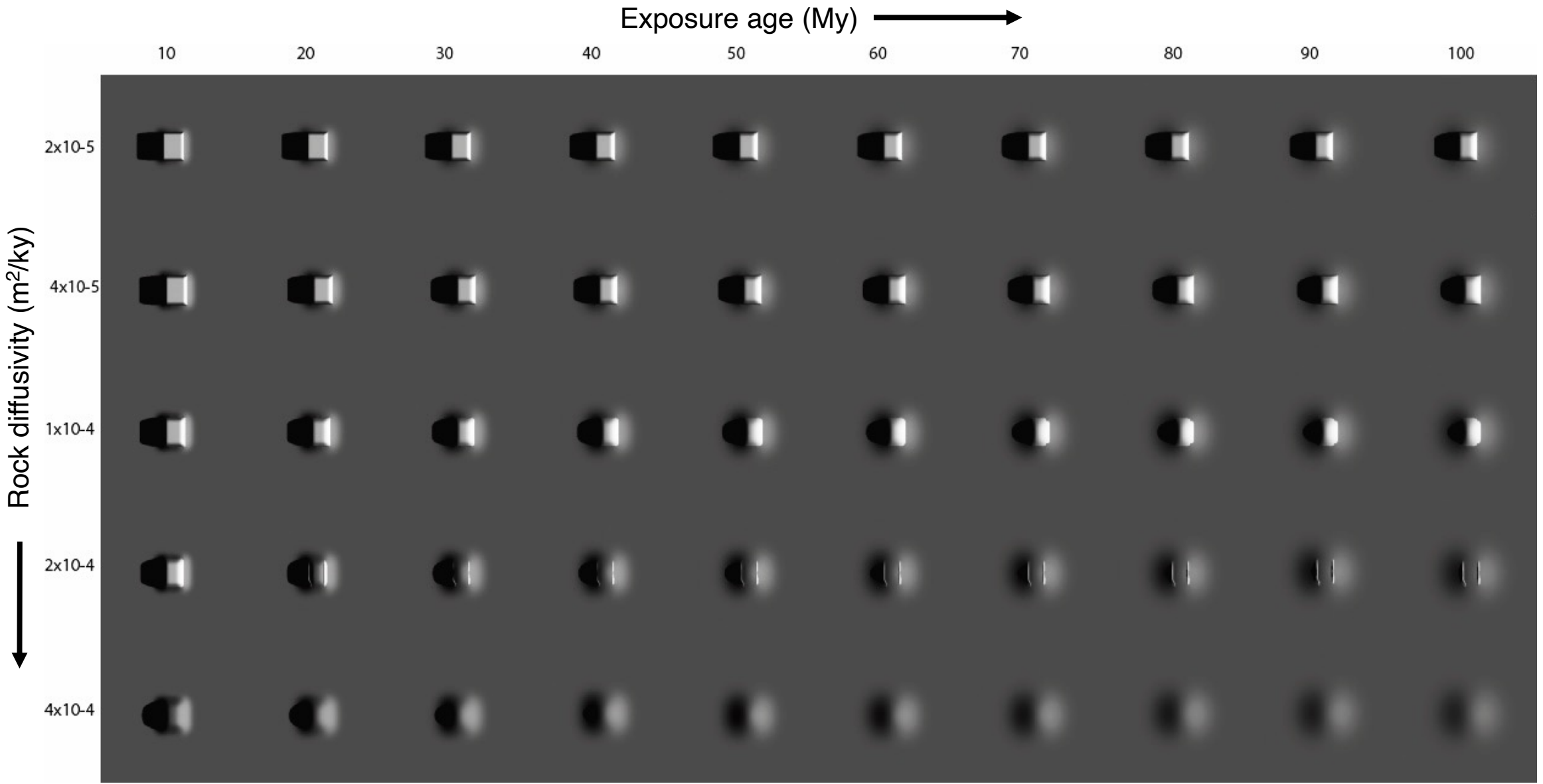

**Figure 12.** Morphological evolution of blocks from 10 to 100 My (left to right) of surface exposure age as seen with artificial orbital images with 60° incidence angle. Block friability increases from top to bottom. Fillet diffusivity coefficient constant at 4E-4 m²/ky. Blocks have a rectangular base with the largest axis 20 m in length.

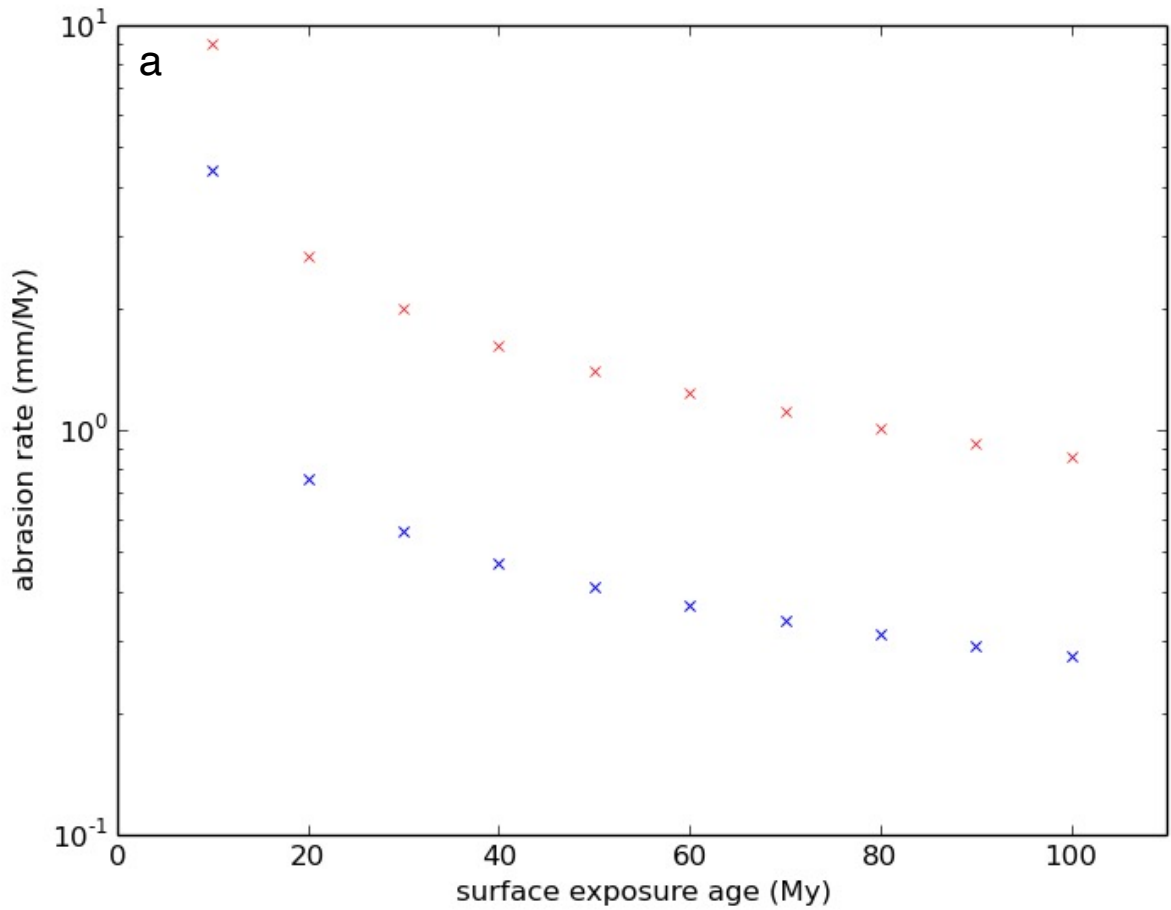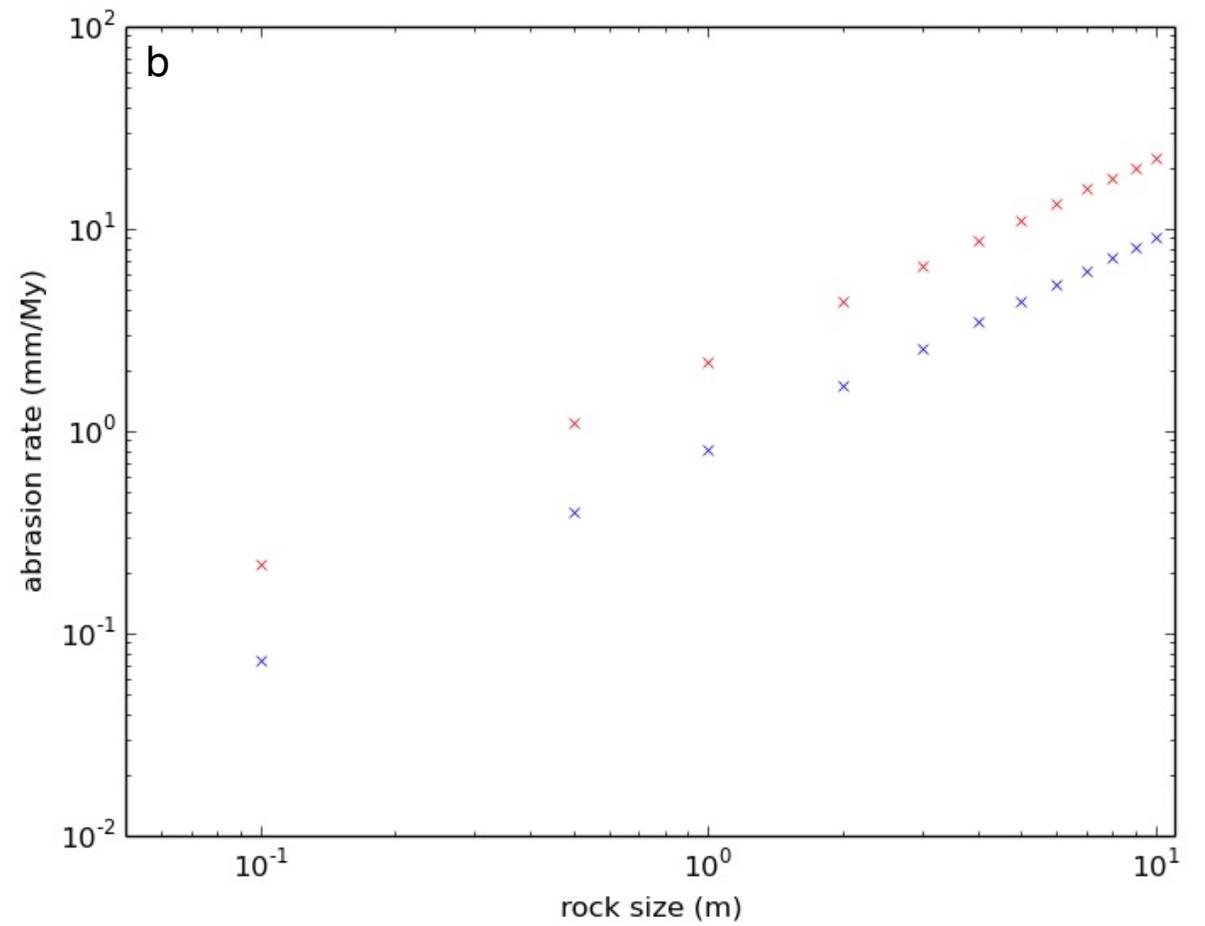

**Figure 13.** (a) Abrasion rate for a one meter wide rock of high (blue) and low (red) strength as a function of surface exposure age. The abrasion rate is calculated each 10 My. (b) Mean abrasion rates after 100 My of surface exposure for a rock of high (blue) and weak (red) strength as a function of rock size. See text for a description of initial rock shape.

| Shown in figure | Block name and location | Width estimate (m) | Age estimate (Myr) | Exposure age and type of estimate | Model block width (m) | Model initial height/width ratio | Model exposure age (My) | Model $K_{rock}$ (m²/kyr) | Model $K_{fillet}$ (m²/kyr) |
|---|---|---|---|---|---|---|---|---|---|
| Fig. 9a, b | Geophone rock, Geophone station, A17 | 4± 0.1 | 106±4 | Sampling: sample 70135 (Arvidson et al., 1975; Drozd et al., 1977) | 5 | 0.54 | 100 | 1.0E-06 | 3.0E-05 |
| Fig. 9c, d | 1005, near B3 station, A14 | 2± 0.1 | 26 | Inferred: Cone crater formation (Arvidson et al., 1975) | 2 | 0.54 | 30 | 2.0E-06 | 6.0E-06 |
| Fig. 9e, f | Large Mound, near ALSEP station, A12 | 2± 0.1 | 50 or 303±18 | Sampling: sample 12008 or 12021 | 2 | 0.54 | 300 | 1.0E-06 | 6.0E-06 |
| Fig. 9g, h | 1006, near B3 station, A14 | 2± 0.1 | 26 | Inferred: Cone crater formation (Arvidson et al., 1975) | 2 | 0.54 | 190 | 2.0E-06 | 6.0E-06 |
| Fig. 11 g,m | 24.27 S / 64.07 W | 25±1 | 47±14 | Byrgius A (Morota et al., 2009) | 20 | 0.54 | 50 | 7.0E-05 | 4.0E-04 |
| Fig. 11 h,n | 24.23 S / 64.02 W | 18±1 | 47±14 | Byrgius A (Morota et al., 2009) | 20 | 0.54 | 50 | 1.0E-04 | 4.0E-04 |

**Table 1.** Measured properties and model parameters for blocks shown in Figure 9 and 11. All blocks are modelled with a square base. The age of Byrgius A crater is presented in Morota et al. (2009) using superimposed crater size-frequency distributions and the chronology function of Neukum (1984) and Neukum and Ivanov (1994).

| Parameter | "Young, high strength" | "Old, high strength" | "Young, low strength" | "Old, low strength" |
|---|---|---|---|---|
| $h_{rock}$ | $r_{rock}*1.08$ | $r_{rock}*1.08$ | $r_{rock}*1.08$ | $r_{rock}*0.87$ |
| $h_{fillet}$ | $h_{rock}*0.25$ | $h_{rock}*0.25$ | $h_{rock}*0.55$ | $h_{rock}*0.55$ |
| $\alpha_{rock}$ | 0.05 | 0.10 | 0.17 | 0.17 |
| $\alpha_{fillet}$ | 1.1 | 0.15 | 0.4 | 0.1 |
| s | 0 | 2 | 1.7 | 2 |

**Table 2.** Parameter values for equations (2) and (3) to describe the topography of abraded rocks with fillet. The names of the sets are qualitative. These parameters are valid to model rocks in cm to m scale.

| Reference | Rock size | Single particle abrasion (mm/My) | Approach |
|---|---|---|---|
| Shoemaker (1971) | 1-10 cm diameter | 1.4–2.1 | Numerical modeling, constant flux over the entire mare history (3.9 Gy) |
| Shoemaker (1971) | 1-10 cm diameter | 0.041–0.062 | Numerical modeling, best estimate for the recent erosion rate (last few My). |
| Gault et al. (1972) | 10 cm diameter | 1.8 | Numerical modeling |
| Ashworth and McDonnell (1973) | 10 cm diameter | 0.1–0.01 | - |
| Neukum (1973) | unspecified dimension | 1 | Numerical modeling |
| Hörz et al. (1974) | 6 cm diameter, crystalline | 0.4–0.6 | Numerical modeling |
| Crozaz et al. (1971) | ~9 cm wide (e.g, crystalline sample 12063) | 0.3–1 | Analysis of apollo samples |
| This study | 10 cm wide, high strength | 0.07 | Numerical modeling. Mean value after 100 My. |
| This study | 10 cm wide, low strength | 0.2 | Numerical modeling. Mean value after 100 My. |

**Table 3.** Summary of abrasion rates reported in the literature and in this study based on numerical modeling and returned apollo samples.